\documentclass[12pt]{mn2e}
\usepackage[dvips]{graphicx}
\usepackage[dvips]{rotating}
\newcommand{\oiii}{[O{\sc iii}]}

\title[Quasar emission lines, radio structures and radio unification]
{Quasar emission lines, radio structures and radio unification}
\author[Jackson \& Browne]
{Neal Jackson$^{1}$, I.W.A. Browne$^{1}$\\
$^{1}$ Jodrell Bank Centre for Astrophysics, 
School of Physics \& Astronomy, 
University of Manchester,
Turing Building, Oxford Road, Manchester M13 9PL\\
}

\begin{document}
\maketitle
\begin{abstract}
Unified schemes of radio sources, which account for different types of
radio AGN in terms of anisotropic radio and optical emission, together
with different orientations of the ejection axis to the line of sight,
have been invoked for many years. Recently, large samples of optical
quasars, mainly from the Sloan Digital Sky Survey, together with large
radio samples, such as FIRST, have become available. These hold the
promise of providing more stringent tests of unified schemes but,
compared to previous samples, lack high resolution radio
maps. Nevertheless they have been used to investigate unified schemes,
in some cases yielding results which appear inconsistent with such
theories. Here we investigate using simulations how the selection
effects to which such investigations are subject can influence the
conclusions drawn. In particular, we find that the effects of limited
resolution do not allow core-dominated radio sources to be fully
represented in the samples, that the effects of limited sensitivity
systematically exclude some classes of sources and the lack of deep radio data
make it difficult to decide to what extent closely separated radio
sources are associated. Nevertheless, we conclude that
relativistic unified schemes are entirely compatible with the current
observational data. For a sample selected from SDSS and FIRST which
includes weak-cored triples we find
that the equivalent width of the \oiii\ emission line decreases as
core-dominance increases, as expected, and also that core-dominated quasars
are optically brighter than weak-cored quasars.

\end{abstract}


\large

\section{Introduction}

Radio sources in the centres of distant galaxies have both historical
and current significance as beacons which indicate high-energy
processes in high-redshift sources. They are also thought to be
important in the life-cycle of galaxies because they may regulate the
processes of star-formation by periodic ejection of gas from the
central regions of the galaxy (e.g. Bower et al. 2006; Croton et
al. 2006; Best et al.  2006; Best \& Heckman 2012; Cattaneo et
al. 2009). Much effort has been devoted to studying radio source
populations in order to understand the physical processes in, and
space distributions of, the different types of active galactic nucleus
(AGN) which host them. For most radio sources with fluxes of more than
a few hundreds of $\mu$Jy, jet emission associated with an AGN is likely
to be the dominant contribution to the radio flux density. A
transition occurs at this level (Richards et al. 2000; Muxlow et
al. 1999, Padovani et al. 2009); in fainter radio quasars, examples of
which can now be detected to intrinsic flux densities of $\sim
1\mu$Jy (e.g. Jackson 2011), synchrotron emission associated with
stellar processes takes over (e.g. Padovani et al. 2011).

The structures of radio sources in AGN typically include a core,
coincident with the centre of the host galaxy, and jets, which extend
from the core and terminate in hotspots, which are in turn surrounded
by more diffuse extended emission known as lobes. However, the
detailed structures are very diverse for several reasons. First, the
presence of prominent hotspots is associated with high-power radio
sources, the so-called FRIIs (Fanaroff \& Riley 1974), whereas
lower-power FRI sources show extended jet/lobe structure which becomes
less bright with distance along the jet and do not have prominent
hotspots. Second, the cores of radio sources contain components which
exhibit superluminal motion, implying that ejection is taking place at
relativistic speeds. This affects particularly our view of a source
whose ejection axis is close to the line of sight, because in this
case the core will appear much more dominant because of relativistic
Doppler boosting. We will then see core radio emission which is
stronger than the other extended components of the source (hotspots
and lobes) which do not move at relativistic speeds. This leads to a
powerful selection effect, in that we will observe intrinsically
faint, apparently core-dominated radio sources which are beamed into
flux-limited samples by this relativistic boosting effect. This
suggests that the ratio of the core to extended radio flux,
$R$\footnote{$R$ is usually defined as the ratio of core- to lobe-
flux density in the emitted frame of the object at 5GHz, with the
necessary $K$-correction carried out using a spectral index of 0 for
the core and $-0.75$ for the lobe emission.}, can be used as at least
a crude orientation indicator, with higher values of $R$ corresponding
to jet ejection angles close to the line of sight.
Such ``unified schemes'' of radio sources were developed 30 years ago
(Blandford \& Rees 1974; Orr \& Browne 1982; Kapahi \& Saikia 1982;
Jackson \& Wall 1999). Subsequently they were modified to incorporate
the idea (Peacock 1987, Scheuer 1987, Barthel 1989) that 
quasars represent a preferentially oriented population whose parent
population, closer to the sky plane, consists of radio galaxies. Radio
galaxies, although similar in appearance to FRII lobe-dominated
radio quasars, do not have strong broad emission lines, and this is
postulated to happen because their broad emission-line regions are
hidden behind a torus-shaped obscuring component, in a similar manner
to the presumed distinction between Seyfert 1 and 2 galaxies. The boundary 
between quasars and radio galaxies is thought to correspond to an orientation
of the outflow axis of approximately 45$^{\circ}$ to the line of sight,
based on statistics of the 3C sample.

As part of these developments, it is natural to ask what the properties of 
the different type of radio source would be in other wavebands, and 
whether these can also be explained by the operation of unified schemes. 
A number of correlations between radio properties and optical structure 
were found. For example, the width of the H$\beta$ line is generally 
larger for low-$R$ objects, as might be expected if the line is emitted 
from a disk whose axis lies along the ejection axis (Wills \& Browne 1986). 
Browne \& Murphy (1987, hereafter BM87) developed this model
to incorporate anisotropic optical and X-ray continuum emission, 
postulating that some fraction of the continuum is beamed, or at 
least anisotropic.

Provided that the narrow emission lines (e.g. \oiii) of quasars 
are approximately isotropically emitted, their equivalent width 
should be lower in high-$R$ objects because the optical continuum 
in such objects would be expected to be relatively stronger.
This was indeed found by Jackson et al. (1989) and Jackson \& Browne 
(1991) in a sample of bright radio quasars for which good radio maps 
and optical spectra were available. In this work, \oiii\ line strengths of
core- and lobe-dominated quasars were matched between quasars of similar
extended radio power, to remove possible luminosity effects. In samples
not so matched (Boroson \& Oke 1984; Boroson, Persson \& Oke 1985;
Fine, Jarvis \& Mauch 2011) the lobe-dominated quasars
tend to have more luminous \oiii\, but even if this effect is not
due to a secondary correlation it is smaller than the differences in
equivalent width that we consider here.

An alternative/additional  hypothesis is that the optical 
continuum in quasars is affected by aspect-dependent extinction
(Baker 1997, but see also Fine, Jarvis \& Mauch 2011) which causes 
the optical continuum of low-$R$ objects,
which are seen at a larger angle to the line of sight,
to be relatively obscured compared to high-$R$ objects, or that
the continuum emission source is itself arranged in a flattened
disk-like structure. Risaliti, Salvati \& Marconi (2011) have analysed
the distribution of \oiii\ equivalent widths in the Sloan Digital
Sky Survey (SDSS, Abazajian et al. 2009) quasar
sample (Schneider et al. 2010, Shen et al. 2011) and find that the 
distribution is consistent with that expected from anisotropic 
continuum emission. Modifications of this model can be made to
explain differences in optical spectra as a function of source
power. For example, ``receding-torus'' models (Lawrence 1991;
Simpson 1998; Grimes, Rawlings \& Willott 2004) propose that the
opening angle of obscuring material around the AGN becomes smaller
as the objects become intrinsically less powerful.

Recently Kimball et al. (2011a,b, hereafter K11a,K11b) have analysed a
large sample of radio sources selected from the SDSS quasar sample
(Schneider et al.  2010). Of the original sample of optical quasars,
they isolate a sample of 4714 quasars with significant detections in
the FIRST 1.4-GHz radio survey (Becker et al. 1995). Of these, a
significant minority of 619 objects have radio extents large enough to
be resolved by visual inspection into a core and two lobes (``triple
sources''), and a further 387 into a core coincident with the quasar
and a single other component, assumed to be a lobe (``lobe
sources''). We will adopt the same terminology throughout the paper.
Having flux densities for both a core and lobe(s) allows the $R$
parameter to be determined and, together with the published SDSS
spectra, yields plots of equivalent widths of a range of emission
lines against $R$. No correlations are found which are significant at
3$\sigma$, although in Table 8 of K11b just for the ``lobe'' sources
the anti-correlation between EW\oiii\ and $R$ is significant at the
2.8$\sigma$ ($\sim 0.5\%$) level, thus apparently consistent with the
earlier results by Jackson et al. (1989) and Jackson \& Browne (1991).

In this paper we discuss the expected correlations of emission line
properties with radio structure, first comparing the small, but well
characterized, Jackson et al. (1989) sample with the much larger
K11a sample. We begin in Section 2 by direct
comparison of the data from the original sample with the new data from
the SDSS. In section 3 we describe the approach of using simulations
to quantify selection effects in the sample and try to assess the
degree to which these effects affect the observed correlations. In Section
4 we discuss the distribution of radio structures within the sample. In
Section 5, given our knowledge of the selection effects we use the
surviving correlation to investigate the range of models which are
still compatible with the data by considering the extent to which the
properties of the simulated and real samples are consistent with each
other. Throughout the paper we assume a flat cosmology with
$H_0=70\,$km$\,$s$^{-1}$Mpc$^{-1}$ and $\Omega_{\Lambda}=0.7$.

\section{Original sample}

The original sample of Jackson et al. (1989) consisted exclusively of
bright radio objects, typically having flux densities of few hundred
mJy at GHz frequencies, unlike the K11a sample which
contains objects with flux densities $S_{\rm 1.4GHz}>2$mJy. On the
other hand, high-resolution (normally sub-arcsecond) radio maps were
available in the smaller sample, compared to the much lower resolution
(5-arcsecond) maps available with FIRST.

We can assess the effect of this difference in resolution by examining
the FIRST data for the Jackson et al. sample. Of the 26 sources from
this sample within the FIRST footprint, 13 are found in the DR5
version of the SDSS quasar catalogue, which had less complete sky
coverage than more recent versions of the catalogue but which we use
throughout this paper for compatibility with K11a. Of these, only
five are visible as resolved sources in the FIRST catalogue, one
knotty source (B1217+023), three triples (B1004+130, B1028+313,
B1512+370) and one lobe (B1111+408).  Of the four triple and lobe
sources, the derived $R$ values for three are in very close
agreement. The fourth, B1111+408=3C254, is very discrepant; for this
source, $\log_{10}R$=$-1.8$ according to Jackson et al., but Table 2
of K11a lists this source as having a core flux density of 2004~mJy
and an extended flux density of 1030~mJy. Examination of
high-resolution radio maps reveals that the former is correct, and
that the origin of the discrepancy is that this source has a very
asymmetric structure, with one lobe close to the core (Owen \&
Puschell 1984; Saikia et al., 1990). On the lower resolution FIRST
images, this lobe appears as part of the core, leading to an
overestimate by a factor of $\sim$100 of the $R$ value, thus
incorrectly leading to its classification as a core-dominated radio
source.

\begin{figure}
\includegraphics[width=8cm]{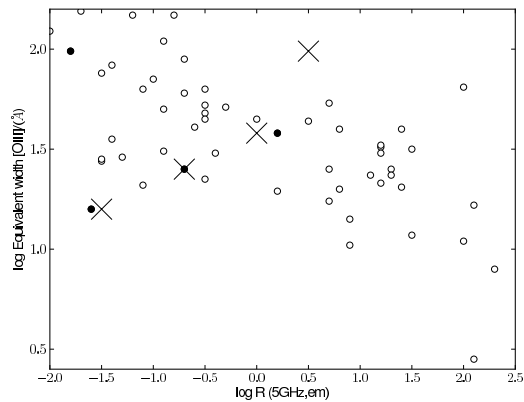}
\caption{Comparison of the \oiii\ equivalent width-log$R$ correlation
for the sample of Jackson et al. (1989) (circles) with the
same sample whose $R$ values have instead been calculated from the
FIRST data (crosses). Only four objects are resolved by the FIRST data, one
of which has an $R$ value affected by blending of the core with one
lobe; these four are plotted as filled circles from higher-resolution
radio data. All points are plotted using the spectroscopic data from
Jackson et al. 3C254 is the anomaly, having moved from core- to
lobe-dominated due to the higher-resolution radio data.}
\label{compare}
\end{figure}

Most of the SDSS quasars in the Jackson et al. sample, however, do not
appear amongst the sources used in the K11a analysis. This is because
they have such small angular extents that they appear as single sources
in FIRST. This unfortunately tends to selectively exclude the more
core-dominated radio sources; of the seven core-dominated radio
sources from the Jackson et al. sample which are both in FIRST and
SDSS-DR5, only one appears as a resolved source in the catalogue of
K11a.  To illustrate this, in Fig.~\ref{compare} we show the
anti-correlation between \oiii\ equivalent width and log$R$ for the
Jackson et al. data and highlight the objects with measured $R$ values
from FIRST. The $R$ values have been corrected to the emitted frame at
5GHz assuming a spectral index of $\alpha=-0.75$, where
$S_{\nu}\propto\nu^{\alpha}$.

\section{Assessing selection effects with simulations}

The analysis in Section 2 indicates that lack of high resolution radio 
maps affects the equivalent width/$R$ correlation.     It is also likely that, 
despite a visual inspection process, some of the K11a ``lobe 
sources'' may be misclassified, again because of limited resolution 
and sensitivity of the radio data. In Section 2 we pointed out that 3C254 has
been misclassified as a high R lobe source simply because the resolution was 
insufficient to separate the second lobe from the adjacent core. Another 
potential for mis-classification happens when one of the lobes of a triple 
source is either too weak or of too low a surface brightness to be detected 
in the FIRST maps. Finally, since the source density is high, some of the 
``lobes'' of lobe sources may be chance associations.    
To assess how important these effects are we have performed some 
simulations. We first look at the effects of resolution and then try 
and quantify the effect of including a significant
number of chance associations in the K11a sample.

\subsection{Selection effects due to FIRST resolution}

The original samples are small and contain mainly very radio-bright
quasars. One possibility is that the fainter quasar
population may be qualitatively different.  The other possibility,
considered here, is that the fainter quasars may
behave in the same way as the brighter ones, but that the use of the
FIRST sample introduces a selection effect in the larger sample due to
the exclusion of core-dominated quasars, primarily because of the
limited angular resolution of FIRST.

We have therefore undertaken studies of a simulated radio population
in order to assess the effect of selection. This has been done using 
the data and methods of Wilman et al. (2006), who simulated a large 
population of radio sources to predict the sky as it would be seen 
by the Square Kilometre Array (SKA). These S$^{3}$ simulations are publicly 
available\footnote{http://s-cubed.physics.ox.ac.uk/s3\_sex. Where 
necessary, as in the case of the FRII sources of which relatively
few examples appear in the S$^{3}$ simulations, we have instead used
the prescription given by Wilman et al., together with the S$^{3}$ 
flux-redshift distribution to generate our own samples. We have checked 
that these agree with the properties of the S$^{3}$ samples, and in
particular that the sample of FRIIs which is generated reproduces
the source counts at low (1.4~GHz) and high (18~GHz) frequencies.}. 
Sources in this sample are chosen from a uniform distribution in
intrinsic size between zero and $(1+z)^{-1.4}$Mpc, and the flux
density and redshift distributions are obtained from measured radio
luminosity functions at 151MHz (Wilman et al. 2006, and references
therein). We assume the 151MHz flux of each source to be unbeamed, 
steep-spectrum emission, and project it to the required frequency of 
1.4GHz using a spectral index of $-$0.75. This gives the extended
flux density, which is the combined flux density of lobes and hotspot.
The core flux density is calculated by multiplying this by $R$, 
using a random angle to the line of sight, $\theta$, to generate 
the $R$ value. Positioning of core, lobe and hotspot components with
respect to each other is carried out assuming that jet axes are
randomly distributed in three dimensions. We modify the prescription
slightly, in order to select only quasars, by choosing only those 
sources with an angle between radio jet axis and the line of 
sight in the range 0$^{\circ}<\theta\leq45^{\circ}$. 




For each object, $R(\theta)$ is given by

\[
R=\frac{1}{2}R_T\left((1-\beta\cos\theta)^{-2}+(1+\beta\cos\theta)^{-2}\right)
\]

\noindent where $R_T$ is the value of $R$ that would be seen in an object 
oriented in the sky plane, and $\beta$ is the ejection velocity as a fraction
of the speed of light. The beaming factor associated with this value 
of $R$ is given by 

\[
g(R)=\frac{1}{2}\left[(1-\mu)^{-2+\alpha}+(1+\mu)^{-2+\alpha}\right]
\]

\noindent where

\[
\mu=\beta\cos\theta=\left[1+\frac{R_T+
\sqrt{8RR_T+R_T^2}}{2R})\right]^{\frac{1}{2}}.
\]

\begin{figure*}
\begin{tabular}{cc}
\includegraphics[width=8cm]{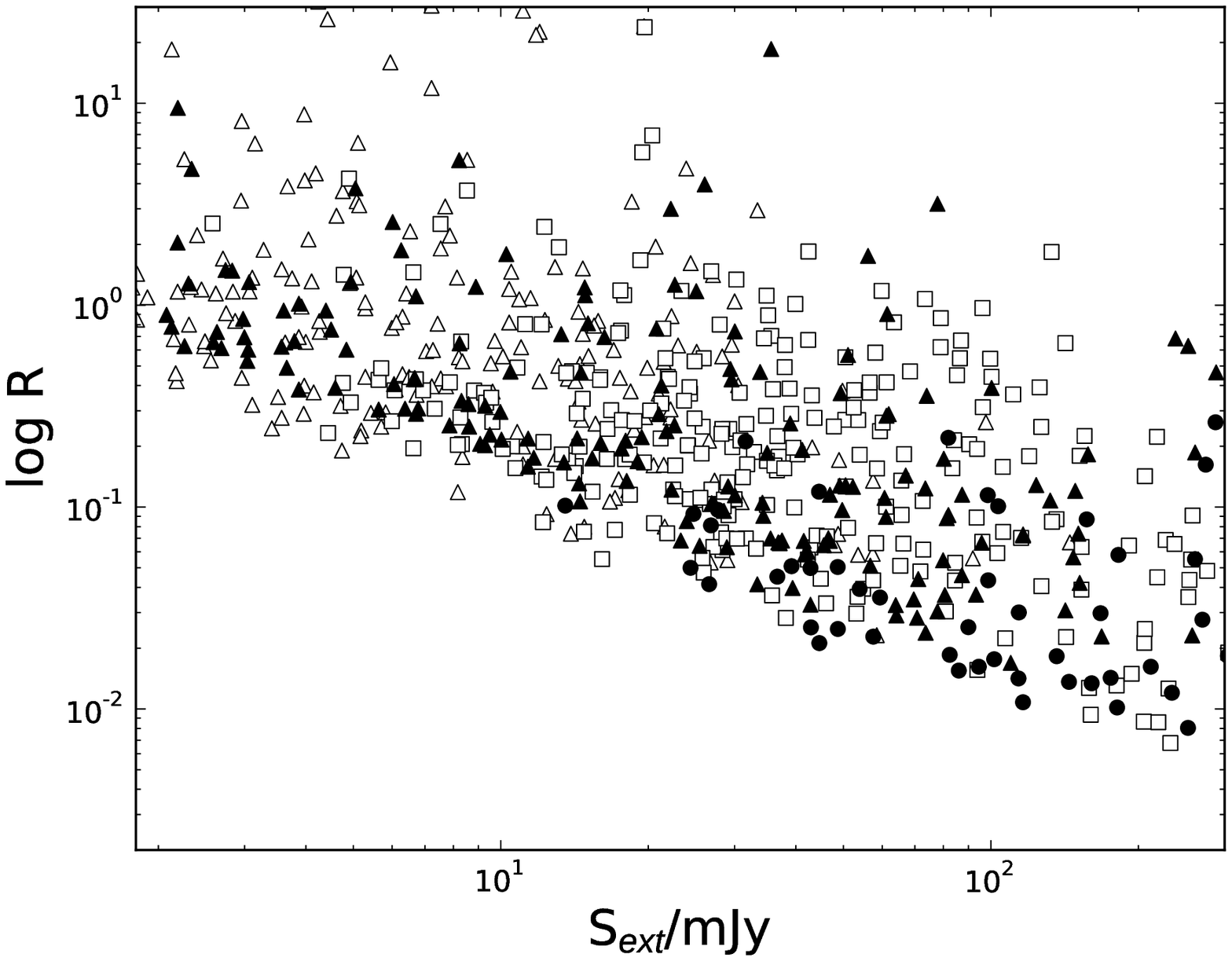}&
\includegraphics[width=8cm]{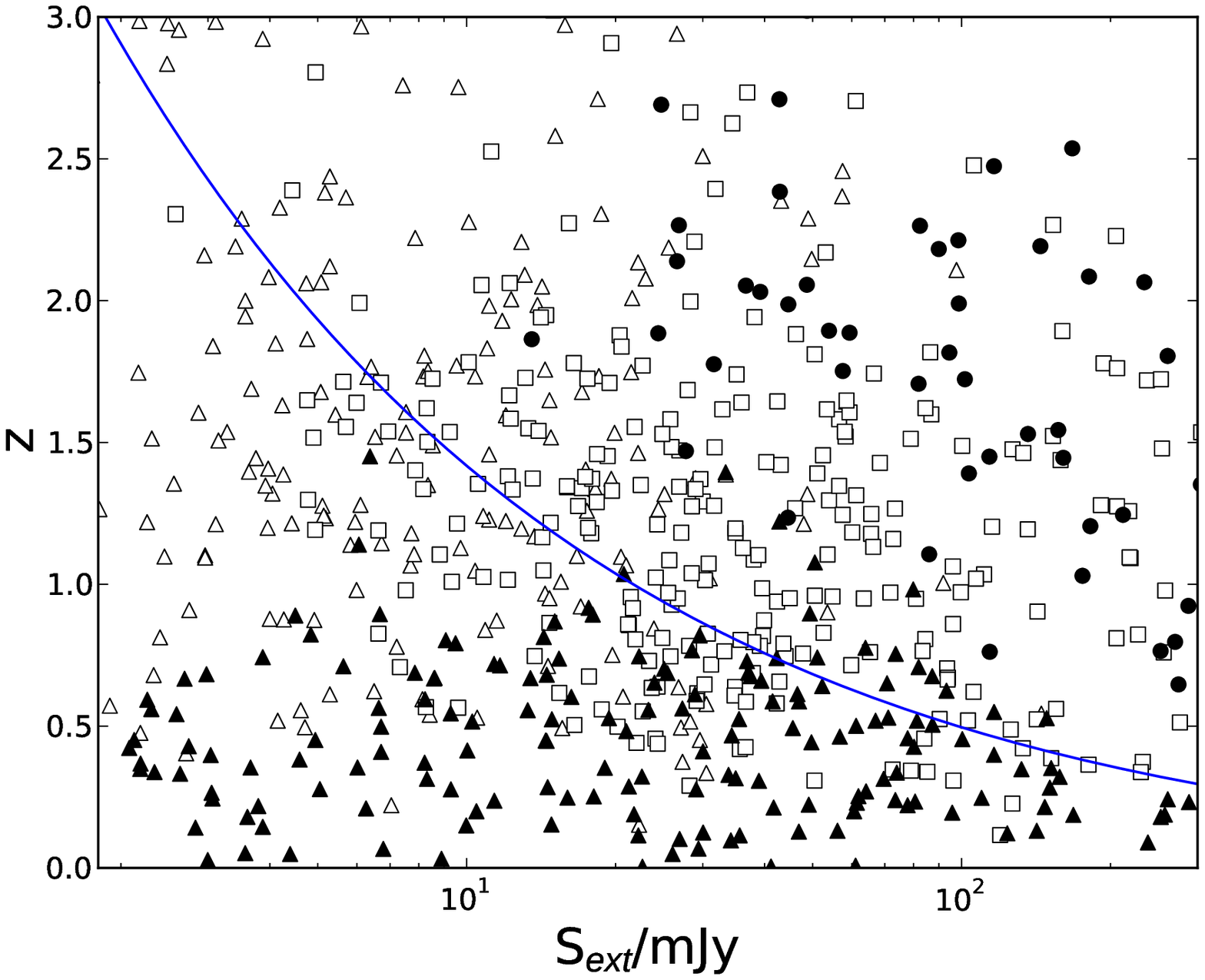}\\
\end{tabular}
\caption{Comparison between the samples of quasars identified from the
SDSS and FIRST surveys with the simulated sample from the SKADS
simulations.  In each case, SKADS FRIs are represented by filled
triangles and FRIIs by filled circles. SDSS/FIRST quasars are
represented by open symbols (only every other quasar has been plotted
for clarity). Triangles represent ``lobe'' sources and squares
``triple'' sources. Left: $\log R$ vs. extended flux density. The
cutoff in the lower left is a result of the limit on core flux,
$S_{\rm core}=RS_{\rm ext}(1+z)^{\alpha}=2$mJy.  Right: redshift vs. extended flux
density; the line represents the boundary in luminosity between FRI
and FRII sources, drawn at a luminosity of 10$^{25}$WHz$^{-1}$sr$^{-1}$.  
Comparison with the redshift distribution indicates
that the SDSS/FIRST quasars are mostly likely to be FRII sources,
although a minority may be FRI. The $R$ value, however, appears to be
generally lower in the simulated FRII sources than those in the
SDSS/FIRST quasar sample.}
\label{frtype}
\end{figure*}

\begin{figure}
\includegraphics[width=10cm]{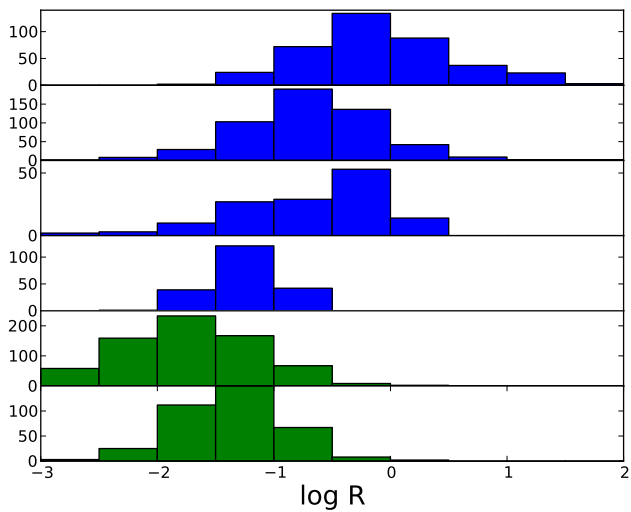}
\caption{Histograms of log $R$ for lobes (top) and triples (second),
both from the data of K11a. Below this are two histograms
constructed from the data presented in Section 4:
weak-core sources with $S_{\rm core}<2$mJy (third; of the original 193
sources, 52 have been rejected by eye as chance associations), 
coreless sources (fourth: note that in this case the $R$ values are
all upper limits, and that of the original 295 sources, 92 have been
rejected by eye). Finally, two histograms are shown from the 
simulated sample from section 3.1, with log$R_T$=-2.8 and
without (fifth) and with (bottom) a cut at 2mJy in the core flux 
density.}
\label{rhist}
\end{figure}

For FRII sources in the Wilman et al. simulations, $\gamma=8$, and the 
distribution in $R_T$ is taken to be a Gaussian distribution with mean 
$10^{-2.8}$ (but we also test other values; see section 4) 
and a dispersion of 0.5 in log space. We make the conventional
assumption that quasars are {\em ipso facto} FRII sources, despite recent
indications that phenomenological boundaries between FRI and FRII may
be slightly blurred (Gendre et al. 2010; Heywood, Blundell \& Rawlings 
2007); this is supported by the finding that our simulated quasars lie
in the same region of $S_{\rm ext}-z$ space as SDSS/FIRST quasars 
(Fig.~\ref{frtype}). 
There is a discrepancy, however, in that the
SDSS/FIRST quasars have $R$ values typically about 0.5--1 dex greater
than those of the SKADS simulation (Fig.~\ref{rhist}). We defer
discussion of this point, but for the present we assume that the
SDSS/FIRST objects resemble standard FRII radio galaxies.

We can then
calculate the expected equivalent width $E$ of the \oiii\ line as
predicted by the BM87 beaming model in which

\[
E = \frac{L_e^p}{A_0L_e^{0.5}+B_0g(R)L_e},
\]

\noindent where $L_e$ is the 5-GHz extended radio luminosity of the source, 
$g(R)$ is the beaming factor corresponding to the particular value of $R$ 
and $A_0$ and $B_0$ are constants, fitted by BM87 as $10^{10.58}$ and 
$10^{-6.78}$ respectively for the case of $p=0.5$. The parameter, $p$ 
quantifies the dependence of the unbeamed optical luminosity on radio 
luminosity and was estimated from the available data by BM87. The first 
term in the denominator represents an isotropic component of the optical 
continuum, and the second 
term represents a relativistically beamed component which dominates
only in cases where $R$, and thus $g(R)$, is large. In practice this
means that noticeable effects on equivalent width are usually only seen
when $R\gg 1$, although the exact boundary
depends on $p$\footnote{We also note that this model was designed for FRII
radio sources, and not for sources of the much lower luminosity of
typical FRI sources: this makes necessary the assumption that in studying 
the SDSS/FIRST sample, we are predominantly dealing with the FRII
population.}.

We have assumed the same functional dependence of the beamed and
unbeamed components as BM87, but have recalculated the $A_0$ and $B_0$
constants. This has been done for two reasons; first, the luminosities
have been corrected for the standard flat-$\Lambda$ cosmology 
rather than the cosmology used by BM87 ($H_0=50$\,km$\,$s$^{-1}$Mpc$^{-1}$, 
$q_0=0$). Second, the hypothesis that quasars are oriented with axes
within 45$^{\circ}$ of the sky plane has been incorporated, as was not
done by BM87. Following Wilman, we used a value of 
10$^{-2.8}$ for $R_T$ at
20cm, with a scatter of 0.5 dex around this value. The recalculated
$A_0$ is always very close to $-10.54$; the value of $B_0$ is ($-8.38,
-7.84,-7.30$) for $R_T$=($-$2.8,$-$2.4,$-$2.0); neither constant is
affected by the value of $p$.


We then assume that objects are recognised in the FIRST survey as
triples only if the core is separated from the nearest hotspot by at
least 5$^{\prime\prime}$ in projection. This assumption has been
investigated by plotting the core-lobe separations for triple sources
in the K11a catalogue. We find a cutoff which begins slightly above
5$^{\prime\prime}$ and tails off to zero at about
4$^{\prime\prime}$. If, however, we generate a sample of sources,
censored by the requirement that they must appear as multiple
components at FIRST resolution, and compare their $R$ distributions
against the lobe and triple sample of K11a, we obtain the
distributions in Fig.~\ref{rhist}. The $R$ values of the K11a triples
are clearly systematically higher than those of the simulation. We
postpone the discussion of why this might be until Section 4.


In summary, if the unified schemes are correct, there are two censorship 
mechanisms which are generated by selection from FIRST. The fact that 
the resolution is limited is bound to exclude some very core-dominated 
sources, whose axes are aligned close to the line of sight, as they will 
typically have projected angular sizes comparable to or less than the 
FIRST resolution. The second selection is that objects whose axes are 
aligned relatively close to 45$^{\circ}$ from the sky plane, the limit at 
which we stop seeing the objects as quasars and start seeing radio galaxies 
instead, will also be missing from the sample due to the cores falling 
below the 2mJy imposed by the FIRST sensitivity cutoff. Thus, if there was 
a real anti-correlation between EW\oiii\ and $R$ as expected in beaming models,
the K11a sample selection with its bias against both high and low $R$ objects 
would mean that such a correlation would not be easy to
see in their data.

%

\subsection{Selection effects arising from chance associations or mis-classifications}


In the previous section we restricted our discussion to triple sources
whose classifications we believe are secure. We note that in the
K11a sample the fraction of sources classified as one-sided
or lobe sources is surprisingly high -- 387 out of a total of 1205 well
resolved sources -- despite their procedure of censoring the sample by
eye to attempt to weed out chance associations. A long established 
feature of radio sources is that nearly all lobes come in pairs, giving 
rise to the term ``double radio sources'' (Moffatt \& Maltby 1961). 
Historically sources with only one lobe are difficult to find (Saikia 
et al., 1990). Hence the high fraction of lobe sources in the K11a 
sample suggests that there may be significant contamination from 
chance associations and/or a significant number classified as lobe sources 
but which in reality are triples with one of the lobes missed.
It is these lobe 
sources that show the strongest anti-correlation between \oiii\ 
equivalent width and $R$ (see Table 8 of K11b) so it is important 
to establish their true nature.

With a source density of 63 per square degree, as is the case for
FIRST sources at $\geq$ 2mJy, we would expect to find several hundred
chance associations giving rise to a ``lobe'' classification, if
sources of separation up to 1$^{\prime}$ from an SDSS quasar are included. Assuming no
clustering, the chance of a random association of a source brighter
than $S$~mJy at a distance less than $D$ arcseconds is

\[
P(\geq S,\leq D) = 3.1\times 10^{-5}\frac{D^2}{S},
\]

\noindent leading to a total of 263 expected ``lobe'' classifications
based on random associations. In reality the above is almost certainly 
a significant underestimate since conclusive evidence for spatial 
clustering in FIRST has been given by Cress et al. (1996), who give
an estimate of clustering signal versus distance for separations of
$>3^{\prime}$ where association between different parts of radio sources
is unlikely to be a problem. If we extrapolate the clustering signal 
from Cress et al. to separations of $<1^{\prime}$, we calculate that
a total of $\sim$400 random associations are likely to be present.  
(In principle, random associations may lead to false classification 
of triple sources, but in practice such cases are likely to be very rare). 


There are 4714 SDSS quasars coincident with FIRST radio sources in the K11a list.
Amongst these we find that there are 1441 objects with one or more
other sources in the field and we deduce from comparing our numbers
with Table 1 of K11a that they have rejected 336
sources because they do not regard them as genuine associations and
have combined them with ``core'' sources in their table. There are 387
sources in their lobe sample. Given that the vast majority of the
619 triples are genuine most of the $\sim 400$ chance
associations are likely to be contained within the lobe sample. There is thus
likely to be contamination at the low tens of percent level of the
K11a lobe sample with random interlopers, both because of this and also
because it can be difficult to decide whether a pair of sources are
genuinely associated, particularly if they are widely separated. 

\begin{figure}
\includegraphics[width=9cm]{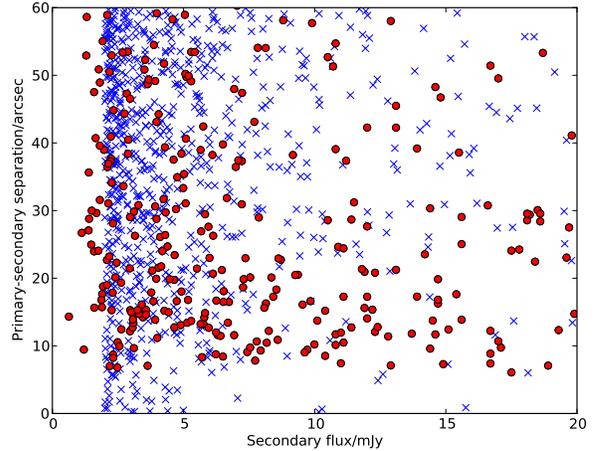}
\caption{The two-image (lobe) sources as a function of separation and
flux of the secondary, both for the observed sample (filled circles) and 
for the same number of random positions within the FIRST survey (crosses).}
\label{interloper}
\end{figure}

We have tried to assess the
effect of random (non-clustered) interlopers in another way, by
picking 4714 random locations within the FIRST survey, a number which
is equal to the total number of radio sources in the K11a 
sample, and testing for sources within a radius of
1$^{\prime}$. The results are shown in Fig.~\ref{interloper} both for
the observed sample and test sources. It appears that sources with
bright secondaries and/or close separations are clearly likely to be
genuinely associated, but since there are many lobe sources in the region
where the density of chance sources is high, the ambiguity in classification
becomes apparent for faint secondaries far from the primary.


Despite the significant residual contamination, the majority of lobe
sources in the K11a sample are likely to be real associations with
SDSS quasars.  What we now discuss is whether they are genuine
one-sided sources or misclassified triples.  This is relevant to the
main theme of the paper, the reality or not of emission line
correlations with radio structure predicted by beaming models, because
we want to know if we can trust the $R$ values of these sources.

In the very simplest models in which radio sources always consist of a
core surrounded by equally spaced and identical lobes, double sources
consisting of a core and one extended lobe should not be seen in any
surveys at any resolution. There are three possible explanations for
the large fraction of sources classified as lobes. First, we could
invoke bend angles which are known to exist in a significant fraction
of these sources (de Vries, Becker \& White 2006) and which would, in
low resolution radio maps, place one of the lobes in projection
against the core and render it unresolvable. We can dispose of this
explanation immediately, as a realistic bend angle of $\sim
15^{\circ}$ can be included in the modelling, and very few sources are
removed by projection of bends in this way. Second, we could be seeing
an intrinsic asymmetry in flux density between two lobes, such that
one of them is located below the surface brightness limit of the FIRST
survey.  Third, the sources could be intrinsically asymmetric in
arm-length ratio, so that one of the lobes is not seen in
low-resolution radio maps. We have already seen one example of this in
the case of 3C254 (see section 2). This can be roughly estimated by
using the projected arm-length asymmetry distribution found for 3C 
quasars by Best et al. (1995). Assuming that the longer arms of the
``triple'' sources from the SDSS-FIRST sample are representative of
the overall population, we can find the distribution of the shorter
arms by taking the longer arm of each such ``triple'' source,
and randomly assigning to it a value of arm-length ratio from the 
3C distribution. This allows the projected distance from the core of 
the closer lobe to be calculated, and in turn allows us to test whether 
the closer lobe would appear at less than 5$^{\prime\prime}$ from the core, 
and thus be blended with it. Such tests
show that the incidence of 3C254-like cases is roughly 4\% for sources
with the core-lobe separation distribution of the SDSS-FIRST quasars. It
is therefore likely that $\sim$20 of the ``lobe'' sources are actually
misaligned triples like 3C254. We also note that, based on examination
of high-resolution images of the 44 3CRR quasars, only 3C254 appears 
misidentified as a lobe. This lends support to the conclusion that
only a few percent of sources are likely to be misidentified in this way.

\begin{figure}
\includegraphics[width=8cm]{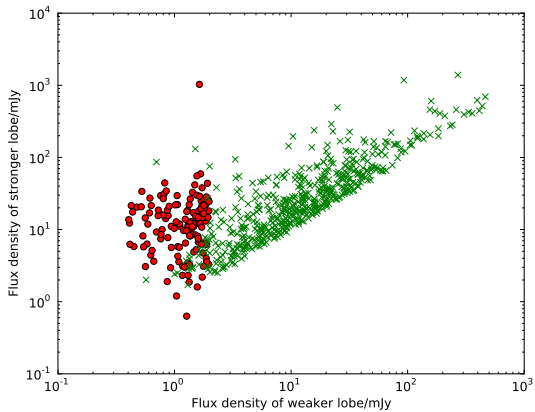}
\caption{Plot of the stronger against the weaker lobe flux for the triple
sources (crosses) and for the lobe (two-component) sources (filled circles).
In the latter case, the flux density of the weaker lobe has been assumed to
be a uniform distribution between 0.4 and 2mJy.}
\label{lobetriple}
\end{figure}

It is relatively easy to investigate whether or not a combination of
censorship by the flux cutoff and intrinsic lobe flux density
asymmetry is the only reason for the apparently large number of lobe
sources. We can do this by plotting the relation between the stronger
and weaker lobe of triple sources (Fig.~\ref{lobetriple}) and then
investigating whether the lobe sources could plausibly lie on this
correlation, allowing for an undetected weaker component at below the
FIRST flux density limit of 2mJy. We have again censored the lobe
sources by removing those close in flux density and separation to a
corresponding interloper from the random sampling of the FIRST
field. On this basis, it is clear that some real triple sources could
be misclassified as lobe sources but it seems implausible that all of
the lobe sources appear as two-component objects simply because of the
FIRST flux density cutoff. There appear to be some strong candidates
for genuine one-sided sources.

\section{Comparison of observed $R$ distribution with simulations}

We now discuss the offset in the $R$ distributions for objects in the
simulations and the the real sample; the sample triples have higher
$R$ values than the simulated triples and overall there seem to too
many high $R$ objects in the whole K11a sample compared to
expectations (Fig.~\ref{rhist}). Though our primary interest is
looking at optical emission line correlations with $R$, understanding
the influences on the observed $R$ distributions is a necessary step.
Initially we will focus on the triples because all these are likely to
be genuine single sources with virtually no contamination with chance
associations.  The implication of the offset in $R$ distributions is
that the selection of the objects for comparison from the simulation
catalogue or the simulation parameters, or both, are slightly
incorrect. We first look at object selection. A significant part of
the explanation for the offset is likely to lie in the requirement
that the SDSS/FIRST sources have a component with flux density $>$2mJy
which is coincident with the SDSS quasar. Many genuine triple quasars
will have radio cores which are weaker than this (Lu et al. 2007). In
order to illustrate this, we have performed a similar analysis to Lu
et al. 2007, using the entire SDSS DR5 sample and correlating it
afresh with the FIRST catalogue in order to identify these missing
triples.  In this correlation, we have demanded that there should be
{\em no} FIRST component within 2$^{\prime\prime}$ of the SDSS quasar;
that there should be exactly two components between 2$^{\prime\prime}$
and 60$^{\prime\prime}$ of the quasar; and that the vectors from the
source to these two components should form an angle of $>90^{\circ}$,
as would be expected from two lobes of a classical double source. This
procedure yields 295 sources, and inspection of the FIRST images
reveals that around 50\% of these are clearly classical double radio
sources whose cores are below the FIRST detection limit. A significant
minority of the remainder are likely to be chance associations. An
examination by eye suggests that there are $\sim$90 of these, but this
number is highly uncertain because of the lack of resolution and
sensitivity in many of the maps. In addition, if we search for
all coreless sources with more than two components between 2$^{\prime\prime}$
and 60$^{\prime\prime}$ from the quasar, we obtain a further 408 sources.
Again examination by eye suggests that a substantial fraction of these
are classical double radio sources centred on the quasar. We do not
include these in the subsequent analysis, but caution that our population
of ``coreless'' sources is likely to be a significant underestimate.

It therefore appears that the 
number of ``coreless'' triple sources is a significant fraction of the
number of triple sources with a detected core. Unsurprisingly, these
``coreless'' triples have lower values of $R$, all of which are upper
limits (Fig.~\ref{rhist}), making the overall FIRST sample, when these
are added in, more comparable in $R$ distribution to the S3-SEX
simulated sample. If the coreless sources with $\geq$3 FIRST components
are included, this is likely to remove the discrepancy altogether.
In addition there are 193 sources with detected
cores $<$2mJy and which have other FIRST radio emission within
1$^{\prime}$. Again, the majority of these are clearly real; an
examination by eye suggests that 52/193 are chance associations, but
again this estimate is highly uncertain.

\begin{figure}
\includegraphics[width=8cm]{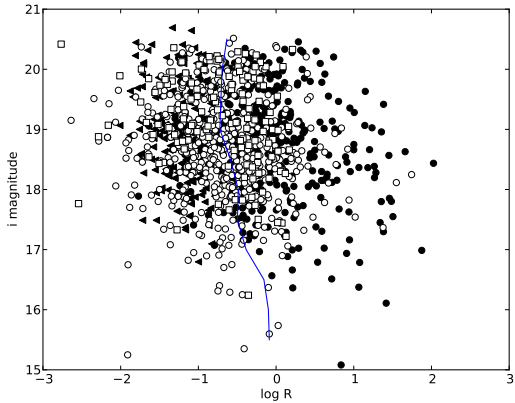}
\caption{The SDSS-$i$ magnitude plotted against log~$R$ for the 
sample. Triple sources are represented by unfilled circles and lobe 
sources with filled circles except that all weak, $<$2mJy core sources
are represented by squares. Coreless sources are represented by
leftward-pointing triangles: note that these are all upper limits in $R$.
The line represents the median value of log $R$ for different
bins of width 1 magnitude, and shows a variation $R$ of 
approximately a factor 3 between faint and bright quasars.}
\label{optmag}
\end{figure}

Although the coreless triples have a similar $R$ distribution to the
simulated sources, this does not account for the observed excess of
high $R$ sources amongst the lobe sources compared to the
simulations. There is another observational selection effect we need
to discuss before deciding if it is necessary to slightly modify the
simulation parameters. This is the effect of having a restriction on the
optical magnitudes of the quasars included in the SDSS quasar
sample. The selection of quasar candidates for optical spectroscopic
follow up is quite complex (see Schneider et al, 2010) but the
majority of objects followed up have i magnitudes brighter than 19.1,
though some fainter objects having colours suggesting that they might
be high redshift quasars have also been followed up. The relevance of a
magnitude limit is that there is an empirical anti-correlation between
$R$ (or equivalently radio spectral index) and optical magnitude
(e.g. Browne \& Wright, 1985). We illustrate this for the SDSS sample
in Fig.~\ref{optmag}. The fact that not all faint quasar
candidates were followed up has introduced a bias against low $R$ objects.

Even when all the selection effects are taken into account it is
likely that the $R$ distribution of the simulations will need to be
modified slightly in order to reproduce the tail of triple and lobe
sources with high $R$. One way to bring the simulation
in line with the observations is to increase $R_T$, as this is
not well constrained by observations. We therefore redo the 
simulations increasing $R_T$ to $10^{-2.4}$ and $10^{-2.0}$ from the value 
of $10^{-2.8}$ suggested by Wilman et al. (2006).  This then increases
all the $R$ values in the histogram by the same factor and makes the
observed and simulated $R$ distributions more
compatible\footnote{Alternatively we could use a distribution in
$\gamma$, which was suggested by BM87 in order to give compatibility
with measured $R$ distributions and source sizes. In practice this
matters less than ensuring that the combination of beaming parameters
used gives a roughly similar distribution in $R$ to that of the
SDSS/FIRST quasars.}. We then re-generate a sample of simulated
objects with an increased value of $R_T$, and appropriately
recalculated values of $A_0$ and $B_0$,
for subsequent comparisons. 

An alternative possibility, which can be investigated by further
high-resolution observations, would be to increase the assumed source 
sizes at low flux density levels so that the censoring by resolution 
has less effect. This would then bring high-$R$ objects back into the
sample which are currently removed by the angular size limit imposed
by the FIRST survey. Such a procedure may be part of a solution, although
the correlation space between linear size, redshift, radio power and 
spectral index is complicated and not yet completely understood (Ker 
et al. 2012).  


\section{Narrow-line equivalent width correlations with $R$}

We now combine the results of the previous sections to investigate the
equivalent width correlations with $R$, including both the resolution 
effects (which affect the global selection of all objects) 
and interloper effects (which affect the ``lobe'' sample).

\subsection{Beaming models}

\begin{figure*}
\includegraphics[width=15cm]{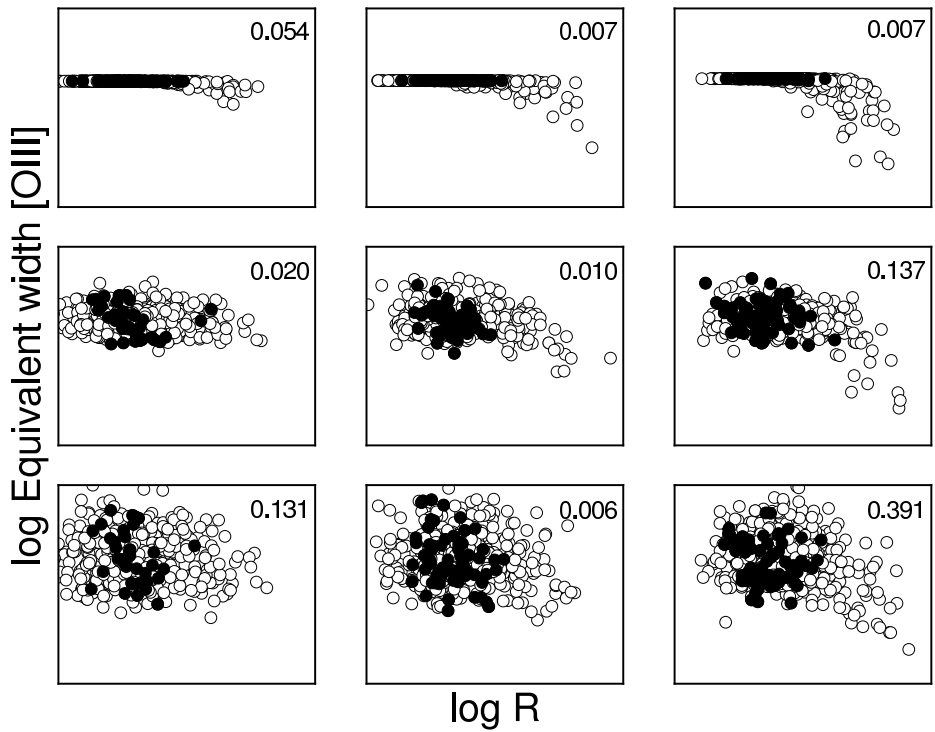}
\caption{Equivalent width of \oiii\ vs. core dominance parameter $R$
(calculated at 20cm) for the full simulated sample, calculated for
$p=0.5$ (top panels), $p=0.75$ (middle) and $p=1.0$ (bottom).  The
left, middle and right panels assume that $\log R_T=-2.8$, $-2.4$ and
$-2.0$, respectively. Open circles represent the full sample. Filled
circles represent those objects which would be visible as observed
(resolved) sources on the FIRST survey images. In most cases, some
objects are not in the observed sample either because they are too
core-dominated, and do not appear as resolved sources because they are
too small, or because they are too lobe-dominated, and the core flux
density falls below the detection threshold. In all plots, the range
of log$R$ is from $-3.0$ to 3.0, the logarithm of the \oiii\
equivalent width is on an arbitrary scale with a range of 1.9 dex, and
the number at the top right represents the probability of the observed
correlation occurring by chance in the censored sample. This
varies considerably between successive random realisations, because it
is usually dominated by a few core-dominated radio sources.}
\label{ewr}
\end{figure*}

Simulations based on the models described earlier allow us to make the
plot of $E$ vs. $R$ expected for both the full uncensored sample, and
for the sample as it would be if selected by the requirement that the
sources should be resolved by FIRST. The results are shown in
Fig.~\ref{ewr} for three values of $p$, 0.5, 0.75 and 1.0; and for the
three values of $R_T$ considered. It is clear that in all cases,
objects at high $R$ and low $E$ are excluded by the FIRST resolution
cut, making the observed correlation less significant than it would
otherwise be.  The degree to which this happens depends on the exact
parameters assumed for beaming model.  A large beaming factor $g(R)$
is required for the beamed optical component to be noticeable, and in
this case the effect on line equivalent width/$R$ correlation is
produced by the few high-$R$ objects which survive the K11a
selection method.

If we look at Fig.~\ref{ewr} in detail we see that the degree of
correlation expected depends strongly on $p$. For $p=0.5$, we expect
all objects in which the beamed component is negligible to have the
same value for the \oiii\ equivalent width; this case corresponds to
the top row of Fig.~\ref{ewr}. In this case the apparent strong
correlations (2-tailed probability $\sim$0.01-0.05) are dominated by
one or two core-dominated quasars. For the other values of $p$ we
expect to see a correlation, the magnitude of which is dependent on
the scatter in $L_e$ and also on the number of beamed objects
present. The latter quantity increases strongly as $R_T$ increases
\footnote{The fraction of flat-spectrum quasars in a flux-limited
sample, for a given Lorentz factor and redshift, varies approximately 
as $R_T^{\delta}$, where $\delta$ is the slope of the integral source 
counts ($N(>S)\propto S^{-\delta}$; Murphy 1988).} 
If we allow both $p$ and $R_T$ to vary over plausible ranges, we
predict a range of correlations which range in significance from
0.006-0.4 in 2-tailed probability.  As discussed previously, because
they reproduce better the range of $R$ seen in the K11a samples we
prefer the values corresponding to higher $R_T$ (right-hand column of
Fig.~\ref{ewr}).  Even here, however, there is a range of allowed
correlations in the censored sample, because of the dependence of the
beaming model on $p$.

\begin{figure}
\includegraphics[width=8cm]{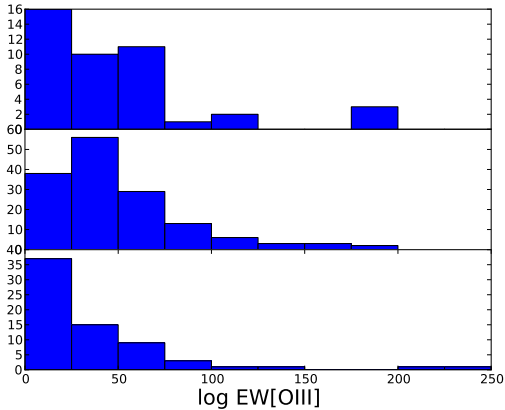}
\caption{The observed distribution of \oiii\ equivalent widths for 
coreless double sources (top), triples (middle) and lobes (bottom).}
\label{ohist}
\end{figure}

What do we actually learn when we compare the observational results
with those of the simulations? We have a combination of a sample of
triples, which are mostly likely to be inclined at a relatively large
angle to the line of sight (because of the selection imposed by the
resolution cut) and in which significant effects of beaming would not
be visible, and a sample of lobe sources. The lobe sources, being most
likely relatively strong cored and weak lobed triples in which one
lobe is not detected (see below), are likely to be inclined closer to
the line of sight, and thus have high R values, but suffer from
interlopers and a few objects like 3C254 at some level. We also have
our own "double sample" (see Section 3.1) containing objects where the
quasar lies roughly between two morphologically recognizable lobes
with but with no detectable radio core. We do not know the individual
$R$ values of the doubles but on average they must be low $R$ objects.
The first thing we do is to compare the \oiii\ equivalent width
distributions of the doubles, triples and lobe sources (see
Fig.~\ref{ohist}). We see that the results are consistent with the
general predictions of beaming (and disk orientation) models with
the doubles, which have low $R$ values, having the highest average
\oiii\ equivalent widths, the triples lower on average and the lobe
sources lowest.

\begin{figure*}
\begin{tabular}{cc}
\includegraphics[width=8cm]{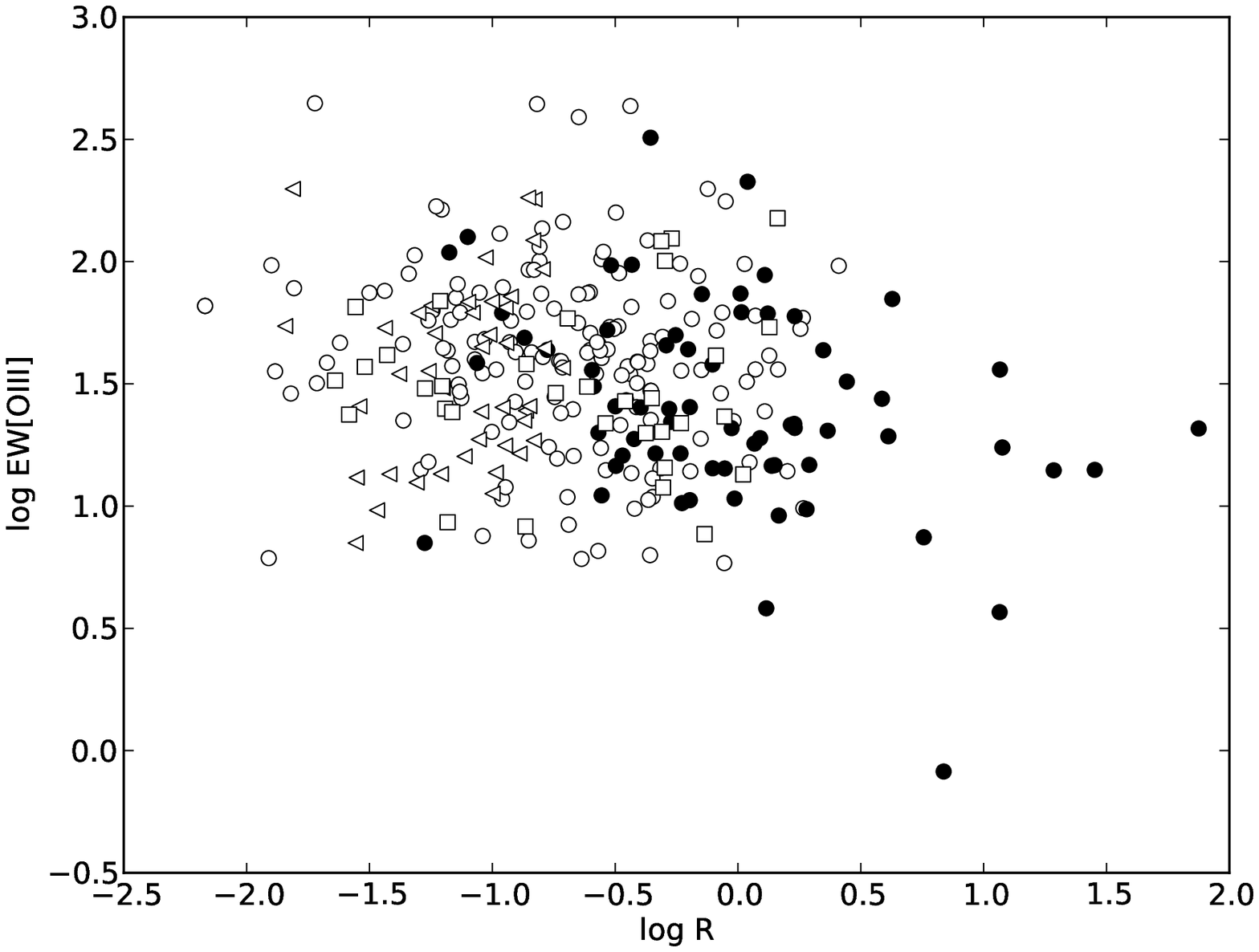}&
\includegraphics[width=8cm]{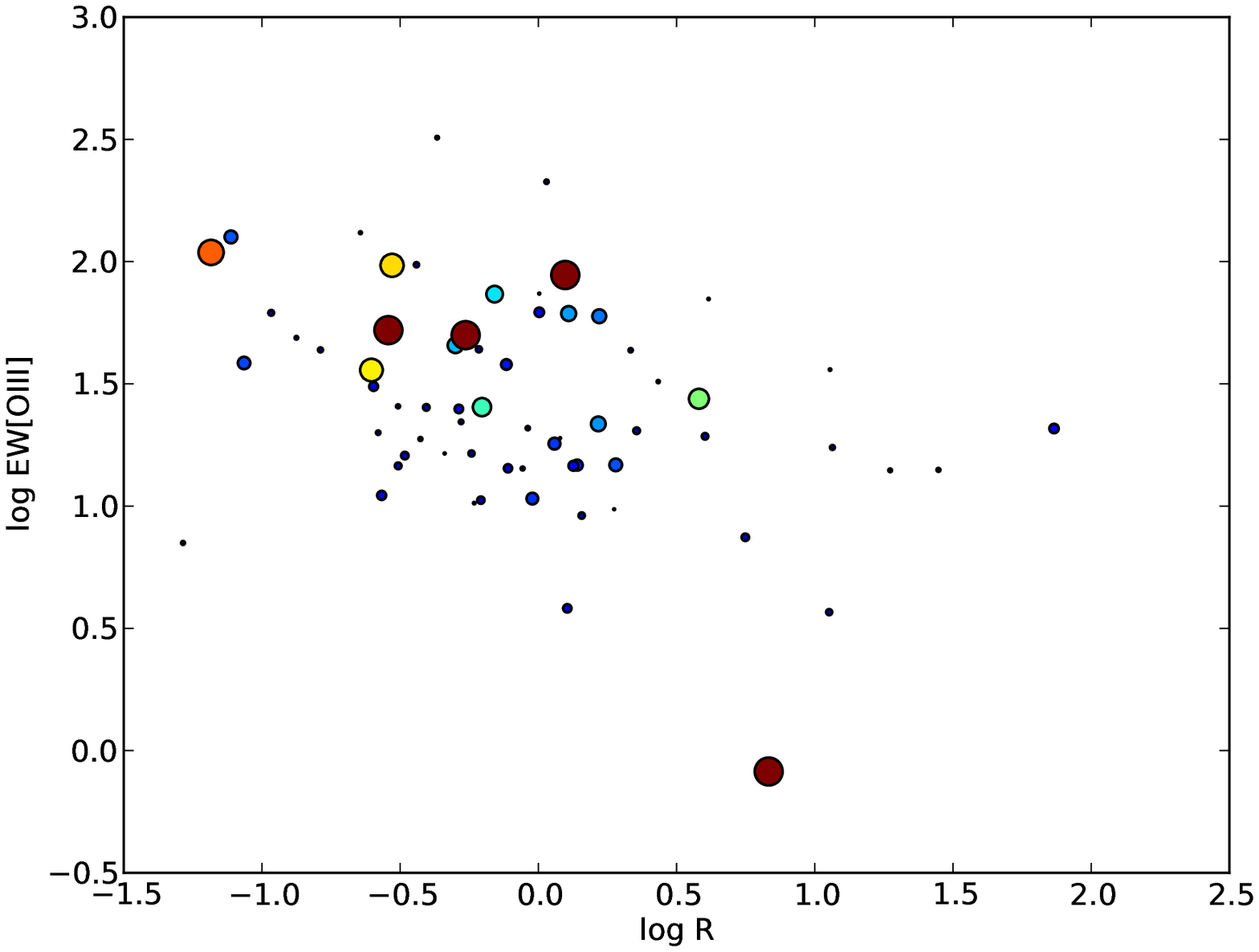}\\
\end{tabular}
\caption{Left panel: Correlations of the logarithm of equivalent 
width of \oiii\ with log $R$ for the triple (unfilled circles) and lobe 
(filled circles) sample, together with sources with weak, $<$2mJy cores
(squares) and coreless sources (leftward-pointing triangles: note that
these are all upper limits in $R$). Coreless and weak-core sources have
been censored by eye (see text) for possible chance associations.
Right panel: the same correlation, but for the lobe
sample and with objects plotted with 
symbol sizes related to the probability of them being a chance 
association (smaller symbols have a higher probability). 
\label{ewr2}
}
\end{figure*}

We can also look at the K11a samples of triples and the lobe samples
individually and in combination (see Fig.~\ref{ewr2}). For the triple
sample alone, the Pearson correlation coefficient is $-$0.13 (0.11
2-tailed probability of occurring by chance). For the lobe sample
alone, the correlation, as in K11b, is significant at approximately
the 0.5\% level. We note, however, that in those of the lobe sources
where the ``lobe'' is just a chance association, the source for which the
equivalent width is measured will by definition be compact and thus in a
beaming model more likely to be seen at a smaller angle to the line of
sight (and hence have a lower value of $E-R$) than more extended sources.
In Fig.~\ref{ewr2} we re-plot the $E-R$ diagram, but with symbol sizes
which are related to the probability of the ``lobe''-source being a
real source rather than a chance association, where closer
associations with a brighter secondary have a higher probability of
being real. Here, the correlation appears to persist among the objects
that are fairly certain to be real associations, although the numbers
become small.

For the combined sample of coreless doubles, triples and lobe sources
the correlation is significant at the 0.12\% level, using the ASURV
software (Isobe, Feigelson \& Nelson 1986) and the generalised Kendall 
$\tau$ test to take account of upper limits on $R$, strongly supporting 
the idea of some general orientation-dependent optical continuum emission. 
Comparing these results in detail with the range
of beaming models described above, we see that most beaming models
can be made to agree with the current data. What is clear is that the
parameters of the FIRST survey are far from ideal for exploring
correlations between emission line properties and radio
structures. Further progress can be made by higher-resolution radio
surveys, as will be made in the far future by the SKA, and in the
nearer future by LOFAR.

\subsection{Disk orientation models}

\begin{figure}
\includegraphics[width=8cm]{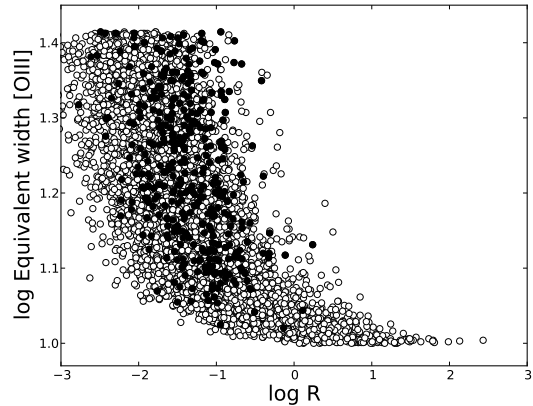}
\caption{Predicted dependence of the \oiii\  against log $R$ for 
a simple disk model. The open circles 
represent all of the sample; the filled circles are those which survive 
the censoring of the FIRST 5-arcsecond resolution.}
\label{disk}
\end{figure}

The equivalent width of emission lines can vary with orientation
either because the the optical continuum emission has a non-thermal
component that is relativistically beamed or because the thermal
emission from a hot, flattened accretion disk will depend on viewing
angle. Such a disk orientation dependence can arise either from
varying inclination angle of an optically thick, geometrically thin
disk (Netzer 1985, 1987) or orientation-dependent obscuration (Baker
1997). The latter model is difficult to quantify but for disk
orientation models, the dependence of line equivalent width on $R$
should be controlled by the variation of the optical continuum flux
with the orientation angle to the line of sight, $\theta$.  The
equivalent width should therefore be proportional to $\sec\theta$,
assuming that no more subtle effects are operating (such as
intrinsically stronger line emission relative to continuum in more
luminous objects). This dependence is shown for the simulated sample
in Fig.~\ref{disk}, together with the objects that survive the
censoring of the FIRST resolution. In this case, we have assumed that
quasars only consist of objects with $\theta<45^{\circ}$, the
remainder of objects being identified with radio galaxies. This
implies that although an anti-correlation should exist, the range of
equivalent widths of \oiii\ expected is not very large (c.f. the
real data shown in Fig.~\ref{ewr2}),
especially when the very core-dominated objects have been removed. In
this case, the value of $R_T$ can again be changed in order to better
fit the range of $R$ seen in the SDSS/FIRST censored sample, but this
does not affect the basic argument.

The Pearson coefficient of the censored simulated sample from
Fig.~\ref{disk} is $-$0.56. As previously suggested, however, the
correlation expected from the triple sample would be expected to be
weaker, due to the restricted range of $\theta$ present in this
sample. As with the beaming models, the expected distribution of
line equivalent widths {\em vs.} $R$ is broadly consistent with the
observed data.

\section{Conclusions}

A dependence of line equivalent width on radio source properties is
expected in beaming models and is potentially an important way to
constrain the model parameters. The currently available radio data do
not enable these parameters to be tightly constrained.  In uncensored
simulated samples a clear anti-correlation is expected, something which
was observed in the original Jackson \& Browne sample. Kimball et
al. have used much larger numbers of objects than used by Jackson \&
Browne in the expectation that these would better constrain beaming
models. However from inspection of Fig.~\ref{ewr} 
we can see that, if we instead restrict
ourselves to data censored by the resolution of the FIRST survey, that
this trend is predicted to decrease or even disappear altogether. This
is primarily because the high-$R$ objects which dominate the trend are
only present in very limited numbers due to sample selection
effects. The actual data, even with selection effects, do show the
expected anti-correlation. By selecting our own sample of doubles from
SDSS and FIRST and adding to the K11a samples of triples and lobe
source we increase the range of $R$ sampled at the low $R$ end of the
distribution. When we do this we can see a systematic trend of
decreasing average equivalent widths with increasing $R$ as expected
for beaming and disk orientation models.

More generally, it is likely that simple dual-population/beaming models and
databases (Jackson \& Wall 1999), even when augmented by the inclusion
of radio-quiet quasars (Jarvis \& Rawlings 2004, Wilman et al. 2006)
may not fully represent the physical processes taking place in radio
sources. One symptom of this is our failure to reproduce the $R$ 
distribution in the censored SDSS/FIRST quasar sample. 
Understanding them will require simultaneous understanding of
orientation and beaming effects, together with the processes of jet
emission, interaction with the external medium and redshift evolution.
This in turn requires the collection of mostly or completely identified samples
of radio sources - not just quasars - with good radio information. A
number of surveys are available at low flux density levels (e.g.
ATESP, Prandoni et al. 2006; TOOT, Vardoulaki et al. 2010; CENSORS,
Best et al. 2003), but the combination of
large numbers, complete optical information and high-quality radio maps
is not yet available. New high-resolution radio instruments, including
MeerKAT, JVLA and e-MERLIN, and eventually the SKA, will be very important
in achieving this goal.

\section*{Acknowledgments}

This research was supported in part by the National Science Foundation
under grant NSF PHY11-25915. We thank Amy Kimball, Patrick Leahy and an
anonymous referee for useful comments on the paper.

\section*{References}
\normalsize

\parindent 0mm

Abazajian K.N., et al., 2009, ApJS, 182, 543

Baker J.C., 1997, MNRAS, 286, 23

Barthel P.D., 1989, ApJ, 336, 606

Becker R.H., White R.L., Helfand D.J., 1995, ApJ, 450, 559 

Best P.N., Arts J.N., R\"ottgering H.J.A., Rengelink R., Brookes M.H.,
Wall J., 2003, MNRAS, 346, 627

Best P.N., Kaiser C.R.,, Heckman T.M., Kauffmann G., 2006, MNRAS, 368, L67

Best P.N., Heckman T.M., 2012, MNRAS, 421, 1569

Blandford R.D., Rees M.J., 1974, MNRAS, 169, 395

Boroson T.A., Oke J.B., 1984, ApJ, 281, 535

Boroson T.A., Persson S.E., Oke J.B., 1985, ApJ, 293, 120

Bower R.G., Benson A.J., Malbon R., Helly J.C., Frenk C.S.,

Baugh C.M., Cole S., Lacey C.G., 2006, MNRAS, 370, 645

Browne I.W.A., Wright A.E., 1985, MNRAS, 213, 97 

Browne I.W.A., Murphy D.W., 1987, MNRAS, 226, 601 (BM87)

Cattaneo, A., et al., 2009, Nature, 460, 213

Cress, C.M., Helfand, D.J., Becker, R.H., Gregg, M.D., White, R.L., 1996, ApJ, 473, 7

Croton D., et al., 2006, MNRAS, 365, 11

Fanaroff B.L., Riley J.M., 1974, MNRAS, 167, 31P 

Fine S., Jarvis M.J., Mauch T., 2011, MNRAS, 412, 213

Gendre M.A., Best P.N., Wall J.V., 2010, MNRAS, 404, 1719

Grimes J.A., Rawlings S., Willott C.J., 2004, MNRAS, 349, 503

Heywood I., Blundell K., Rawlings S., 2007, MNRAS, 381, 1093

Isobe T., Feigelson E.D., Nelson P.I., 1986, ApJ, 306, 490

Jackson C.A., Wall J.V., 1999, MNRAS, 304, 160

Jackson N., Browne I.W.A., 1991, MNRAS, 250, 422 

Jackson N., Browne I.W.A., 1991, MNRAS, 250, 414 

Jackson N., 2011, ApJ, 739, L28

Jackson N., Browne I.W.A., Murphy D.W., Saikia D.J., 1989, Natur, 338, 485 

Jackson N., Browne I.W.A., Shone D., Lind K.R., 1990, MNRAS, 244, 750

Jarvis M., Rawlings S., 2004, New Astr. Rev., 48, 1173

Kapahi V.K., Saikia D.J., 1982, JAA, 3, 465

Ker L.M., Best P.N., Rigby E.E., R\"ottgering J.H.A., Gendre M.A., 2012,
MNRAS 420, 2644

Kimball A.E., Ivezi{\'c} {\v Z}., Wiita 
P.J., Schneider D.P., 2011, AJ, 142, 143 

Kimball A.E., Ivezi{\'c} {\v Z}., Wiita 
P.J., Schneider D.P., 2011, AJ, 141,182

Lawrence A., 1991, MNRAS, 252, 586

Lu Y., Wang T, Zhou H., Wu J., 2007, AJ, 133, 1615

Moffatt A.T., Maltby, P., 1961. Nat, 191, 453

Murphy D.W., 1988, PhD thesis, University of Manchester

Muxlow T.W.B., Wilkinson P.N., Richards A.M.S., Kellermann,  K.I.,
Richards E.A., Garrett M.A., 1999, NewAR, 43, 623

Netzer H., 1985, MNRAS, 216, 63

Netzer H., 1987, MNRAS, 225, 55

Orr M.J.L., Browne I.W.A., 1982, MNRAS, 200, 1067 

Owen F.N., Puschell J.J., 1984, AJ, 89, 932 

Padovani P., et al., 2011, ApJ, 740, 20

Padovani P. et al., 2009, ApJ, 694. 235

Peacock J.A., 1987, in ``Astrophysical jets and their engines'' Proc. 
NATO Advanced Study Institute, Erice, Italy, Sept. 17-25, 1986. 
Dordrecht: Reidel p.185.

Richards E.A., 2000, ApJ, 53, 611

Risaliti G., Salvati M., Marconi A., 2011, MNRAS, 411, 2223

Saikia D.J., et al., 1990, MNRAS, 245, 408

Schneider D.P., et al., 2010, AJ, 139, 2360 

Scheuer P.A.G., 1987. in ``Superluminal Radio Sources'', p.104, eds.
Zensus J.A., Pearson T.J., Cambridge University Press

Shen Y, et al., 2011, ApJS, 194, 45

Simpson C., 1998, MNRAS, 297, L39

Thomasson, P., Saikia D.J., Muxlow, T.W.B.,  2006, MNRAS, 372, 1607 

Vardoulaki E., et al., 2010, MNRAS, 401, 1709

de Vries W., Becker R., White R.L., 2006, AJ, 131, 666

Wills B.J., Browne I.W.A., 1986, ApJ, 302, 56

Wilman R.J., et al., 2008, MNRAS, 388, 1335
\end{document}